\documentclass[aps,twocolumn,prd,showpacs,showkeys,preprintnumbers,superscriptaddress,nobibnotes,floatfix,longbibliography,notitlepage,nofootinbib]{revtex4-1}

\pdfoutput=1
\usepackage{amsmath}
\usepackage{amsfonts}
\usepackage{amssymb}
\usepackage{mathrsfs}
\usepackage{graphicx}
\usepackage{xcolor}
\usepackage{slashed}
 
\usepackage{hyperref}
\usepackage{multirow}
\usepackage{upgreek}
\usepackage[capitalise]{cleveref}

\newcommand{\Fig}{Fig.}

\newcommand{\vmin}{v_{\rm min}}

\newcommand{\ud}{\text{d}}
\newcommand{\sumint}{\int\kern-1.2em\sum}

\begin{document}

\preprint{KEK-QUP-2022-0002, KEK-TH-2450, KEK-Cosmo-0296, IPMU22-0048}

\title{Halo-Independent Dark Matter Electron Scattering Analysis with In-Medium Effects}

\author{Muping Chen}
\email{mpchen@physics.ucla.edu}
\affiliation{Department of Physics and Astronomy, UCLA,\\
475 Portola Plaza, Los Angeles, CA 90095, USA}

\author{Graciela B. Gelmini}
\email{gelmini@physics.ucla.edu}
\affiliation{Department of Physics and Astronomy, UCLA,\\
475 Portola Plaza, Los Angeles, CA 90095, USA}

\author{Volodymyr Takhistov} \email{vtakhist@post.kek.jp}
\affiliation{International Center for Quantum-field Measurement Systems for Studies of the Universe and Particles (QUP, WPI),
High Energy Accelerator Research Organization (KEK), Oho 1-1, Tsukuba, Ibaraki 305-0801, Japan}
\affiliation{Theory Center, Institute of Particle and Nuclear Studies (IPNS), High Energy Accelerator Research Organization (KEK), Tsukuba 305-0801, Japan
}
\affiliation{Kavli Institute for the Physics and Mathematics of the Universe (WPI), UTIAS, \\The University of Tokyo, Kashiwa, Chiba 277-8583, Japan}
\date{\today}

\begin{abstract}
Dark matter (DM)-electron scattering is a prime target of a number of direct DM detection experiments and constitutes a promising avenue for exploring interactions of DM in the sub-GeV mass-range, challenging to probe with nuclear recoils.  We extend the recently proposed halo-independent analysis method  for DM-electron scattering, which allows to infer the local DM halo properties without any additional assumptions about them, to include in-medium effects through dielectric functions of the target material. We show that in-medium effects could significantly affect halo-independent analysis response functions for germanium and silicon and thus are essential for proper inference of local DM halo characteristics from direct DM detection data.
\end{abstract}
\maketitle

\section{Introduction}

The nature of predominant constituent of matter in the Universe, dark matter (DM), remains unknown beyond its gravitational interactions. Numerous proposals have been put forth to explore its possible non-gravitational interactions (see e.g. for review~\cite{Bertone:2004pz,Gelmini:2015zpa}). While significant efforts have focused on studying DM consisting of Weakly Interacting Massive
Particles (WIMPs) with typical masses in the GeV to 100 TeV range that often appear in
models that can address the hierarchy problem, a wide range of DM candidates covering orders of magnitude in mass-range are feasible. One well motivated possibility is that of DM consisting of light sub-GeV mass particles, which can appear in variety of models~(e.g.~\cite{Feng:2008ya,Boehm:2003hm,Lin:2011gj,Hooper:2008im,Hochberg:2014dra,Hochberg:2014kqa}).
 
Traditional direct detection searches focus on energies deposited from Galactic halo GeV-mass DM interacting with nucleons (see e.g.~\cite{Gelmini:2018ogy,Akerib:2022ort}). Due to kinematics that puts energy deposited in scatterings with nuclei below experimental thresholds, sub-GeV DM interactions with electrons at low-threshold experiments constitutes a preferred paradigm. A broad range of studies have explored DM-electron interactions in experiments based on noble gases and semiconductors, and a slew of experimental search proposals have been put forth (e.g.~\cite{Hochberg:2015pha,Hochberg:2015fth,Hochberg:2016ntt,Hochberg:2017wce,Derenzo:2016fse,Kurinsky:2019pgb,Griffin:2019mvc,Blanco:2019lrf,Trickle:2019nya,Geilhufe:2019ndy,Hochberg:2019cyy,Coskuner:2019odd,Griffin:2020lgd}). Already exploited as prime target of experimental collaborations such as XENON1T~\cite{XENON:2019gfn,XENON:2020rca} and XENONnT~\cite{Aprile:2022vux}, DAMIC~\cite{DAMIC:2019dcn}, SENSEI~\cite{Barak:2020fql}, 
and SuperCDMS~\cite{Amaral:2020ryn}, 
testing DM-electron interactions is
poised to  be
of importance in future direct DM detection searches~\cite{Essig:2022dfa}. 

DM-electron scattering 
in noble gases, like xenon, involve interactions with individual atoms 
(see e.g.~\cite{Essig:2011nj}). Crystal targets allow to achieve lower detection thresholds at the level of $\sim\mathcal{O}(1)$eV compared $\sim\mathcal{O}(1)$keV of noble gases, due to their band structure. However, lattice many-body effects
complicate the description of DM-electron interactions in crystals. In Ref.~\cite{Graham:2012su,Lee:2015qva} DM-electron scattering in crystals was calculated with semi-analytic approximations for the electron wave functions.
Numerical calculations based on density functional theory (DFT) to obtain the crystal band structure and electron wave functions were developed in Ref.~\cite{Essig:2011nj,Essig:2015cda,Derenzo:2016fse}. Another approach based on DFT was presented in Ref.~\cite{Trickle:2019nya,Griffin:2019mvc}, subsequently extended to combine DFT with semi-analytic approximations to include a broad range of transition states near and further away from the band gap~\cite{Griffin:2021znd}.

Since 1980's~\cite{Ahlen:1987mn} conventional direct DM detection analyses focused on assuming a model of the local
DM velocity distribution and density, obtaining limits on DM mass-reference cross section $(m_{\chi}, \sigma_{\rm ref})$ space for a particular type of DM interaction. On the other hand, halo-independent analyses avoid the uncertainties associated with our knowledge of the local Galactic halo at the small scales relevant for direct detection and allow to instead infer the local DM distribution from signals as well as directly compare results between distinct experiments. While the halo-independent method has been extensively explored for DM-nucleon scattering (see e.g.~\cite{Fox:2010bz,Fox:2010bu,Frandsen:2011gi,Gondolo:2012rs,HerreroGarcia:2012fu,Frandsen:2013cna,DelNobile:2013cta,Bozorgnia:2013hsa,DelNobile:2013cva,DelNobile:2013gba,DelNobile:2014eta,Feldstein:2014gza,Fox:2014kua,Gelmini:2014psa,Cherry:2014wia,DelNobile:2014sja,Scopel:2014kba,Feldstein:2014ufa,Bozorgnia:2014gsa,Blennow:2015oea,DelNobile:2015lxa,Anderson:2015xaa,Blennow:2015gta,Scopel:2015baa,Ferrer:2015bta,Wild:2016myz,Gelmini:2015voa, Gelmini:2016pei,Witte:2017qsy,Gondolo:2017jro,Ibarra:2017mzt,Gelmini:2017aqe,Catena:2018ywo}), halo uncertainties can also significantly impact DM-electron searches~\cite{Maity:2020wic,Radick:2020qip} and only recently
Ref.~\cite{Chen:2021qao} formulated the halo-independent analysis for DM-electron scattering.

\begin{figure*}[t]
    \centering
    \includegraphics[trim={0cm 0cm 0 0},clip,width=.9\columnwidth]{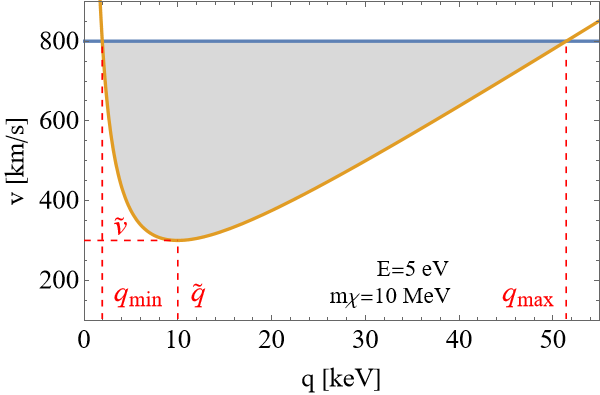}
\hspace{3em}
    \includegraphics[trim={0cm 0cm 0 0},clip,width=.9\columnwidth]{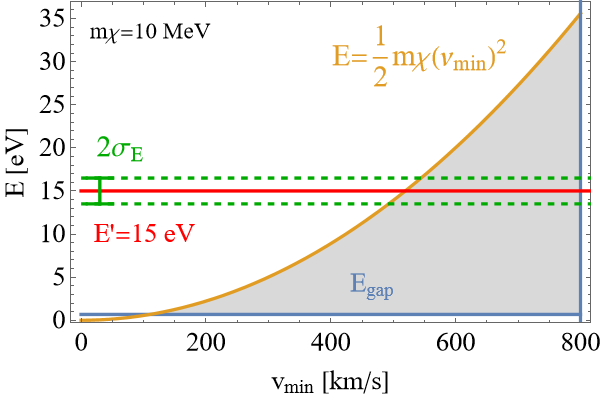}
   \caption{[Left]
   Function $\vmin(q,E)$ (orange line) for $m_{\chi}=10 \,{\rm MeV}$ and $E=5\,{\rm eV}$. Indicated are $\tilde{v}=\sqrt{2E/m_{\chi}}$, The minimum $\vmin$ value corresponding to $\tilde{q}=\sqrt{2m_{\chi}E}$, which separates the left and right $q$ branches, as well as $q_{\rm min}=q_-(v_{\rm max},E)$, and $q_{\rm max}=q_+(v_{\rm max},E)$ are indicated. Here we take the possible maximum DM speed to be $v_{\rm max}=800\,{\rm km/s}$, corresponding to DM bound to the Galaxy. The kinematically allowed region is shown in grey.
  [Right]
  $E$ integration domain in  Eq.~\eqref{eq:responsefunc-plusminus} 
  (grey region) as function of $\vmin$ for $m_{\chi}=10 \,{\rm MeV}$ between the band gap energy (0.67~eV for Ge and 1.1~eV for Si) and the maximum possible recoil energy for a fixed $\vmin$, $E=m_\chi \vmin^2/2$. The  box resolution function we assume is shown for $E'=15$~eV and $\sigma_{\rm E}=1.5\,{\rm eV}$.
   }
    \label{fig:branchesintdo}
\end{figure*}

Collective in-medium effects in condensed matter systems can significantly modify the DM-electron scattering rates, as first noted for a vector mediator (dark photon)~\cite{Hochberg:2015fth} and subsequently for a scalar mediator as well~\cite{Gelmini:2020xir}. These effects could be effectively accounted for through the dielectric function, which is well-studied in a broad range of materials and directly related to the scattering rate~\cite{Hochberg:2021pkt,Knapen:2021run,Knapen:2021bwg} (see also discussion in \cite{Griffin:2021znd}).
  
In this work we build on results of Ref.~\cite{Chen:2021qao} to formulate the halo-independent DM-electron scattering analysis including in-medium effects, based on the dielectric function.

\section{Dark Matter-Electron Scattering}

We first provide an overview of the DM-electron scattering rate, including in-medium effects, as employed in conventional direct DM detection analysis (i.e. in a ``halo-dependent analysis'').

All the relevant in-medium effects are specified via the dielectric function $\epsilon$, which is experimentally well determined for a broad range of materials, as the energy loss function ${\rm Im}[-1/\epsilon(E, \vec{q})]$, where $E$ and $\vec{q}$ are the energy and momentum imparted to the electron.
This is  incorporated in the dynamic structure factor $S$ that describes the rate of creating density
fluctuations in the medium and  is related to the longitudinal dielectric response function $\epsilon = \epsilon_L$ via~\cite{phillips_2012}
\begin{equation}
S(E,\vec{q}) = \dfrac{q^2}{2 \pi \alpha} \dfrac{1}{1 - e^{- \beta E}} {\rm Im}\Big[\dfrac{-1}{\epsilon(E,\vec{q})}\Big]~,
\end{equation}
where $\beta = k_B T$ for temperature $T$ and Boltzmann constant $k_B$, $\alpha$ is the fine structure constant.

The complete expression for the time average (over a full year) DM-electron scattering event rate, in number of counts per unit time per unit mass, including the structure factor $S$ incorporating in-medium effects through the dielectric function $\epsilon$, is then
given by \cite{Knapen:2021bwg}
\begin{align}\label{eq:rate}
	R& =\frac{1}{\rho_T} \frac{\rho_{\chi}}{m_{\chi}} \frac{\sigma_{\rm ref}}{\mu_{\chi e}^2} \frac{\pi}{\alpha} \int \ud^3 v f_\chi(\vec{v})   \int \frac{\ud^3 \vec{q}}{(2\pi)^3} \,q^2   |F_{\rm DM}(q)|^2 \\
 &~\times \int\!\frac{\ud E}{2\pi}\,  \,   \frac{1}{1 - e^{-\beta E}} {\rm Im} \left[ \frac{-1}{\epsilon(E,\vec{q})} \right]\delta\left( E + \frac{q^2}{2 m_{\chi}} - \vec{q} \cdot \vec{v} \right). \notag
\end{align}
Here $\rho_\chi$ is the local DM density, $\rho_T$ is the target density, $\mu_{\chi e} = m_{\chi} m_e/(m_{\chi} + m_e)$ is the DM-electron reduced mass, $F_{\rm DM}(q)$ is the DM-mediator form factor, which depends on the mediator mass, and
$f_\chi(\vec{v})$ is the  time average distribution of DM velocity $\vec{v}$ with respect to the detector, normalized to 1, which in the halo-dependent analysis is typically taken to correspond to that of the Standard Halo Model.

\begin{figure*}[t]
    \centering
    \includegraphics[trim={0cm 1.7cm 0 0},clip,width=\columnwidth]{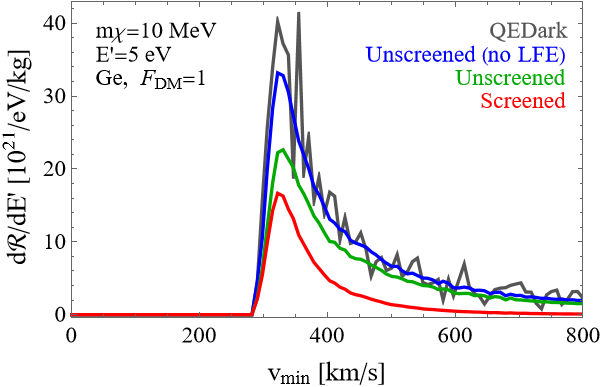}
    \includegraphics[trim={0cm 1.7cm 0 0},clip,width=\columnwidth]{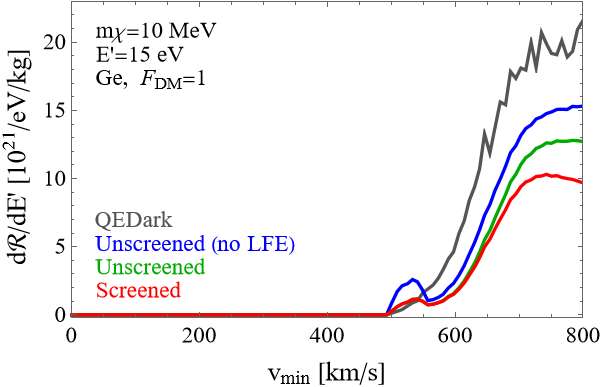}
    \includegraphics[trim={0cm 0cm 0 0},clip,width=\columnwidth]{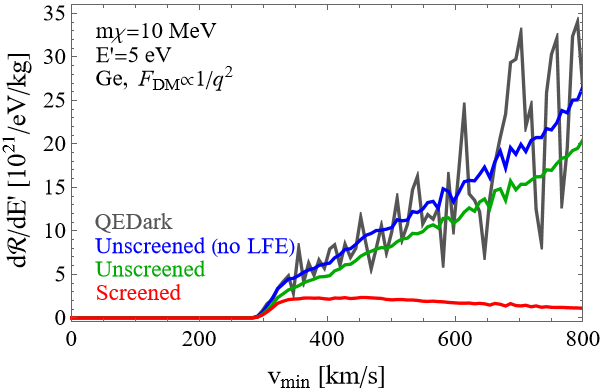}
    \includegraphics[trim={0cm 0cm 0 0},clip,width=\columnwidth]{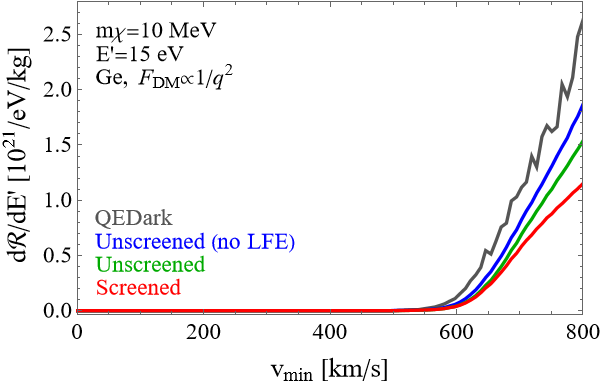}
    \caption{
    Response function in germanium for $\tilde{\eta}(\vmin)$,  $\ud \mathcal{R}(\vmin,E')/\ud E'$,  defined in Eq.~\eqref{eq:responsefunc} 
    calculated with different methods:   \textsf{QEDark}(gray) and three different approaches in \textsf{DarkELF}, namely unscreened without LFE (blue), unscreened with LFE (green) and screened with LFE (red),  for $E'=5$~eV (left panels) and $E'=15$~eV (right panels), with DM-mediator form factor $F_{\rm DM}=1$ (upper panels) and $F_{\rm DM}\sim 1/q^2$ (lower panels). Halo properties can be inferred from data only where $\ud \mathcal{R}/\ud E'\neq 0$.
    }
    \label{fig:dmflux}
\end{figure*}

\begin{figure*}[t]
    \centering
    \includegraphics[trim={0cm 1.7cm 0 0},clip,width=\columnwidth]{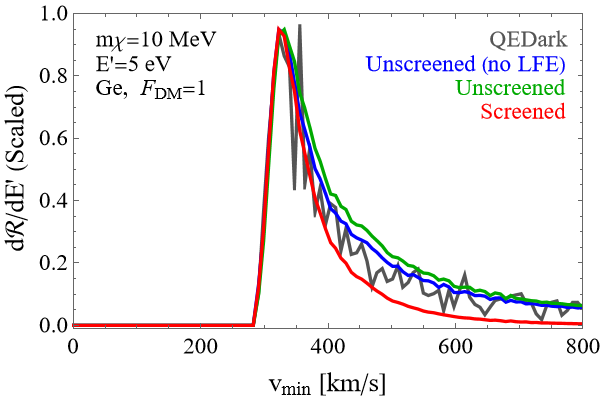}
    \includegraphics[trim={0cm 1.7cm 0 0},clip,width=\columnwidth]{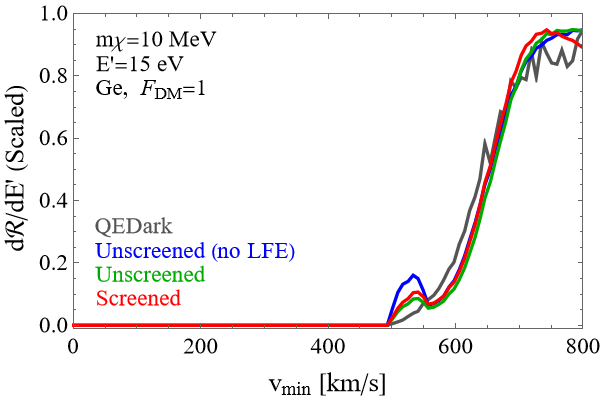}
    \includegraphics[trim={0cm 0cm 0 0},clip,width=\columnwidth]{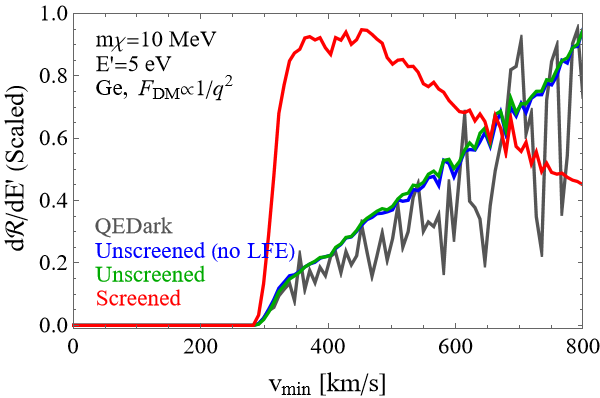}
    \includegraphics[trim={0cm 0cm 0 0},clip,width=\columnwidth]{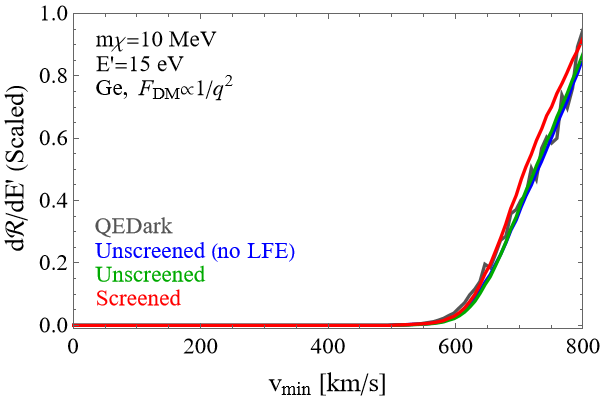}
    \caption{
    Same as Fig.~\ref{fig:dmflux} but the response functions $\ud \mathcal{R}/\ud E'$ are scaled so their maximum is approximately 1, to better show their shape and thus the range of $\vmin$ which each allows to explore when interpreted as window functions.} 
    \label{fig:dmfluxscaled}
\end{figure*}

\begin{figure*}[t]
    \centering
    \includegraphics[trim={0cm 1.7cm 0 0},clip,width=\columnwidth]{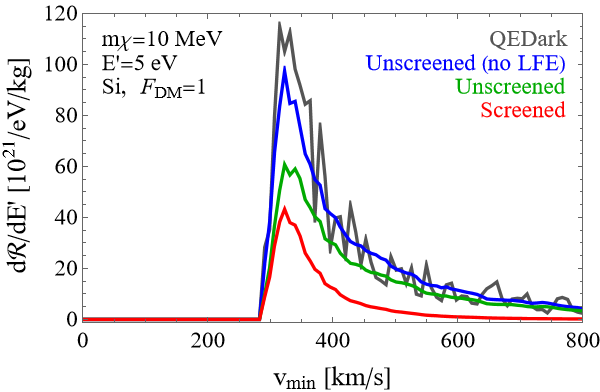}
    \includegraphics[trim={0cm 1.7cm 0 0},clip,width=\columnwidth]{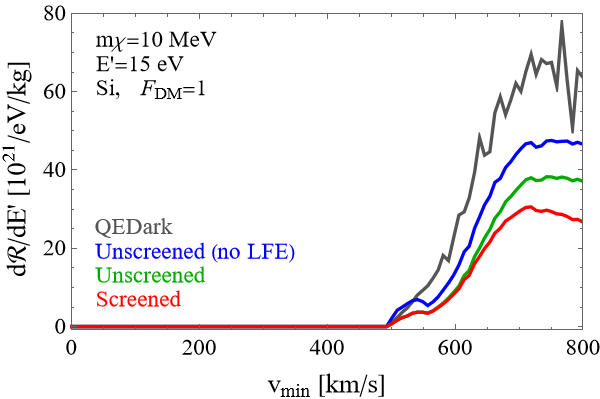}
    \includegraphics[trim={0cm 0cm 0 0},clip,width=\columnwidth]{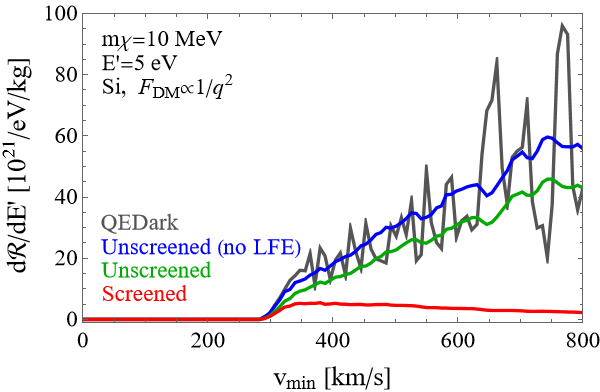}
    \includegraphics[trim={0cm 0cm 0 0},clip,width=\columnwidth]{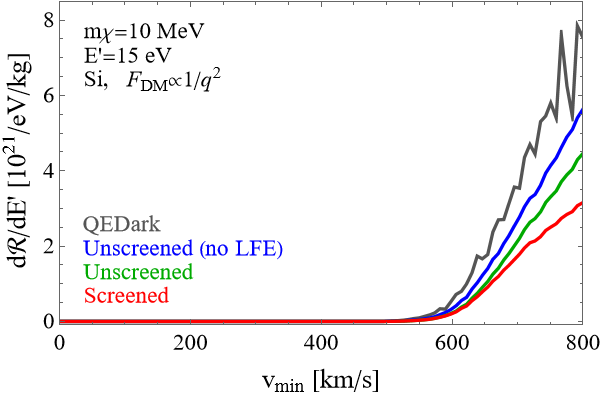}
    \caption{
    Same as Fig.~\ref{fig:dmflux} but for silicon.
    }
    \label{fig:Sidmflux}
\end{figure*}

\begin{figure*}[t]
    \centering
    \includegraphics[trim={0cm 1.7cm 0 0},clip,width=\columnwidth]{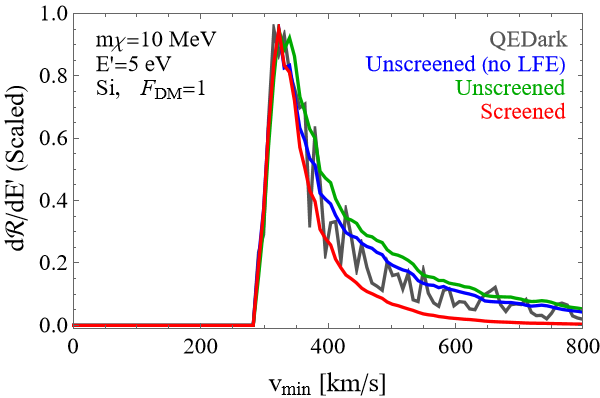}
    \includegraphics[trim={0cm 1.7cm 0 0},clip,width=\columnwidth]{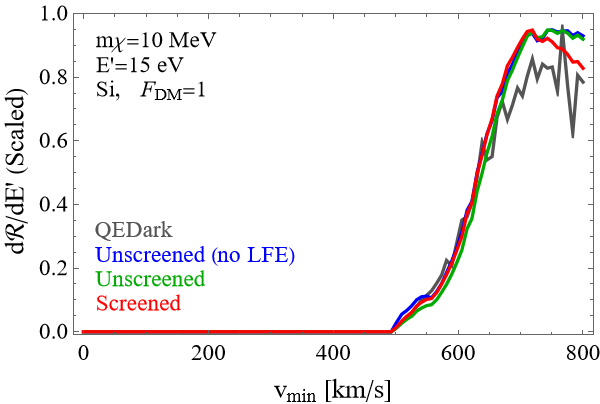}
    \includegraphics[trim={0cm 0cm 0 0},clip,width=\columnwidth]{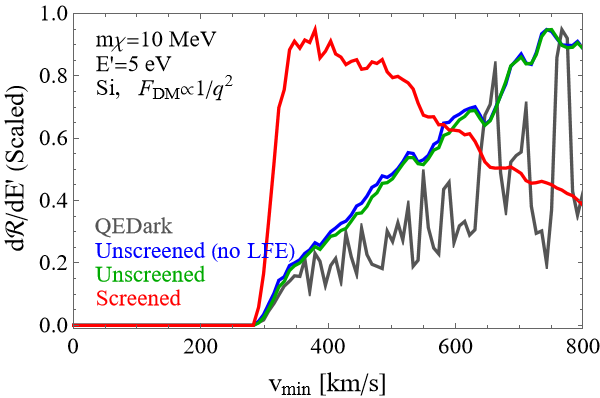}
    \includegraphics[trim={0cm 0cm 0 0},clip,width=\columnwidth]{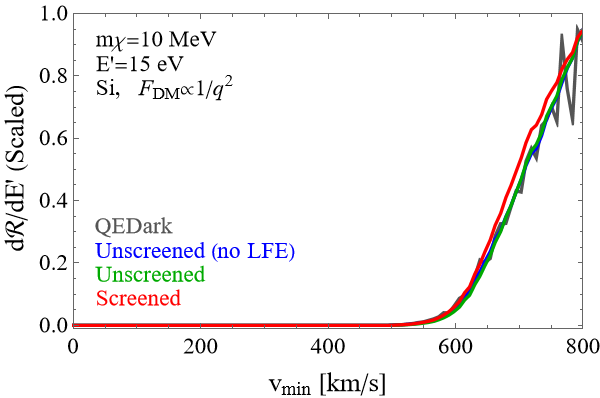}
    \caption{
    Same as Fig.~\ref{fig:dmfluxscaled} but for silicon.} 
    \label{fig:Sidmfluxscaled}
\end{figure*}

\section{Halo-Independent Analysis}

The halo-independent analysis method is based on separating astrophysical quantities contributing to DM scattering rate from the particle physics and experiment-specific quantities contributing to it. The predicted time average  scattering
rate can be written in terms of a function $\tilde{\eta}$ characterizing the local DM halo convoluted with a ``response function'', a kernel that encodes the detector and particle model information as
\begin{equation} \label{eq:haloind}
    \dfrac{\ud R}{\ud E
    }(E) = \int_0^{\infty} \ud v_{\rm min}\, \tilde{\eta}(v_{\rm min}) \dfrac{\ud \mathcal{R}}{\ud E} (v_{\rm min},E)~.
\end{equation}
Here  $\ud R/\ud E$ is the energy differential average rate (per unit time per unit mass), 
$\ud \mathcal{R}/dE$ is the response function for an energy $E$, and the function 
\begin{align}
\label{eq:tilde-eta-t}
\tilde{\eta}(\vmin) 
\equiv&~ \frac{\rho_{\chi} \sigma_\text{ref}}{m_{\chi}}
\int_{v> \vmin}\ud^3 v \, \frac{f_\chi(\vec{v})}{v} \notag\\
=&~ \frac{\rho_{\chi} \sigma_\text{ref}}{m_{\chi}}
\int_{\vmin}^{\infty} \, \ud v \, \frac{F(v)}{v}\ ~, 
\end{align}
 includes all the DM halo dependence of the rate. Here $F(v) \equiv v^2 \int \ud \Omega_v \,f_\chi(\vec{v},t)$ is the speed $v= |\vec{v}|$ distribution. The aim of the halo-independent analysis method is to derive the halo function $\tilde{\eta}$ using direct detection data. Then,  data from different direct detection experiments all detecting DM should produce compatible $\tilde{\eta}$ functions. 

For DM-electron scattering $\vmin$, the minimum speed of the DM particle necessary to produce a recoil energy  $E$ and momentum transfer $\vec{q}$ in a target electron, is
\begin{equation} \label{eq:vmin}
\vmin = \dfrac{E}{q} + \dfrac{q}{2 m_{\chi}}~.
\end{equation}

To bring the rate to the form in Eq.~\eqref{eq:haloind}, we reformulate Eq.~\eqref{eq:rate}.
Using Eq.~\eqref{eq:vmin}, we can  rewrite the delta-function ensuring energy conservation in Eq.~\eqref{eq:rate} as
\begin{equation}
\delta ( E + \frac{q^2}{2 m_{\chi}} - q v \cos \theta_{qv}) = \dfrac{1}{qv} \delta\Big(\cos \theta_{qv} - \dfrac{v_{\rm min}}{v}\Big)~.
\end{equation}
where $\theta_{qv}$ is the angle between vectors $\vec{q}$ and $\vec{v}$. Performing the integration over the solid angle $\Omega_{qv}$ we obtain
\begin{align}\label{eq:rate2}
	R& =~ \frac{1}{\rho_T}   \frac{ 1}{\mu_{\chi e}^2} \frac{\pi}{\alpha}  \dfrac{1}{(2 \pi)^3}   \int dq  \int dE ~  q^3   |F_{DM}(q)|^2 {\rm Im} \left[ \frac{-1}{\epsilon(E,\vec{q})} \right]\notag\\
 &~\times  \frac{1}{1 - e^{-\beta E}}  
 \Big[\int \ud^3 v\,
\frac{\rho_{\chi} \sigma_{\rm ref}}{m_{\chi}} \Theta(v - \vmin) \dfrac{f(\vec{v})}{v} 
 \Big]
 ~, 
\end{align}
where $\Theta$ is the step function. Notice that the term in the square brackets is $\tilde{\eta}(v_{\rm min})$ defined in Eq.~\eqref{eq:tilde-eta-t}.

Taking the derivative of Eq.~\eqref{eq:rate2} with respect to $E$  and performing the change of variables $q$ to 
$\vmin$, we obtain the desired differential DM-scattering rate in terms of $\vmin$,
\begin{align}\label{eq:diffratevmin}
	\dfrac{\ud R_{\pm}}{\ud E} =&~ \frac{1}{\rho_T} \dfrac{1}{8 \pi^2 \mu_{\chi e}^2 \alpha} \int_0^{v_{\rm max}} \ud\vmin\, J_{\pm}(E, \vmin)~  \\
&~\times	q^3(E, \vmin) |F_{\rm DM}(q(E, \vmin)|^2 \notag\\
 &~\times   \frac{1}{1 - e^{-\beta E}} {\rm Im} \left[ \frac{-1}{\epsilon(E,q(\vmin,E))} \right] 
\tilde{\eta}(\vmin)~,\notag
\end{align}
where  $J_{\pm} (E, \vmin) = \partial q_{\pm}/\partial \vmin$ is the Jacobian due to the change of variables.

For a fixed $E$, the momentum $q$ has two solutions $q_{\pm}(\vmin,E)$ as function of $\vmin$. The two $q$ branches meet at the minimum $\vmin$ value $\tilde{v}=\sqrt{2E/m_{\chi}}$, where $q$ takes the value $\tilde{q}= q_{+}(\tilde{v},E)=q_{-}(\tilde{v},E)=\sqrt{2m_{\chi}E}$, as illustrated in the left panel of \Fig~\ref{fig:branchesintdo}. In this figure the blue line horizontal line shows the maximum possible speed $v_{\rm max}$ of a DM particle in Earth's frame which we take to be $v_{\rm max}=800\, {\rm km/s}$. Also indicated are the minimum and maximum values $q$  can take for a given $E$,   $q_{\rm min}=q_{-}(v_{\rm max},E)$ and $q_{\rm min}=q_+(v_{\rm max},E)$. The possible values of the energy $E$ range from the energy gap $E_{\rm min}=E_{\rm gap}$ (0.67~eV for Ge and 1.1~eV for Si)~\cite{experimentgap,Klein:1968}, to $E_{\rm max}=m_{\chi} v_{\rm max}^2/2$,  the maximum kinetic energy the DM particle can have before scattering. 

The total differential rate is then
\begin{equation}
\label{eq:tot=+&-}
    \frac{\ud R}{\ud E}=
    \frac{\ud R_+}{\ud E}-\frac{\ud R_-}{\ud E}~.
\end{equation}
Here, the differential rate of the left branch, $\ud R_- / \ud E$,  carries a negative sign due to the interchange of the lower and the upper limit in the $\vmin$ integral in Eq.~\eqref{eq:diffratevmin}.  

Experiments do not directly measure the recoil energy $E$, but rather a proxy for it that we denote $E^{\prime}$ 
 such as some amount of heat or a number of photoelectrons. We account for this by relating the true differential recoil rate $dR/dE$ to the detected differential recoil rate $dR/dE^{\prime}$ as
\begin{equation} \label{eq:resol}
\dfrac{\ud R}{\ud E^{\prime}} = \varepsilon(E^{\prime}) \int_0^{\infty} \ud E\, G(E^{\prime},E) \dfrac{dR}{dE}~,
\end{equation}
where the $\varepsilon(E^{\prime})$ function accounts for the detector efficiency and $G(E^{\prime},E)$ is the energy resolution function of the experiment. For simplicity, for our figures we assume a box resolution function with width $2 \sigma_E$ centered at $E'$ and $\sigma_E= 0.1 E'$.

Combining Eq.~\eqref{eq:diffratevmin}, Eq.~\eqref{eq:tot=+&-}, and Eq.~\eqref{eq:resol}, we obtain the halo-independent analysis response function $d\mathcal{R}/dE^{\prime}$, 
defined in Eq.~\eqref{eq:haloind} by replacing $E$ by $E'$, as
\begin{equation} \label{eq:responsefunc}
\dfrac{d\mathcal{R}}{dE^{\prime}} (E', \vmin)= \dfrac{d\mathcal{R}_+}{dE^{\prime}}- \dfrac{ d\mathcal{R}_-}{dE^{\prime}} 
\end{equation}
where
\begin{align} \label{eq:responsefunc-plusminus}
	\dfrac{d\mathcal{R}_{\pm}}{dE^{\prime}}& (E', \vmin) = \frac{1}{\rho_T} \dfrac{\varepsilon(E^{\prime})}{8 \pi^2 \mu_{\chi e}^2 \alpha}  \int \ud E\, G(E^{\prime},E)     \\
 &~\times J_{\pm}(E, \vmin)  q_{\pm}^3 (E, \vmin) |F_{\rm DM}(q_{\pm}(E, \vmin)|^2   \notag \\
 &~\times \frac{1}{1 - e^{-\beta E}} {\rm Im} \left[ \frac{-1}{\epsilon(E,q_{\pm}(\vmin,E))} \right]~.\notag
\end{align}

\section{Computation Method}

The computation of the response function  in 
Eq.~\eqref{eq:responsefunc}, requires specifying the integration domain, the energy resolution function, the DM-mediator form factor, and the dielectric function of the specific material. 
 
The right panel of Fig.~\eqref{fig:branchesintdo} shows the integration domain  in $E$ for each fixed $\vmin$ value. It  goes between $E_{\rm gap}$   and $E=m_\chi \vmin^2/2$ (shown as the yellow curve in Fig.~\eqref{fig:branchesintdo}), the maximum possible recoil energy due to a collision of a DM particle moving with speed $\vmin$ (we can see this corresponds to inverting the function $\tilde{v}(E)$ when taking $\tilde{v}=\vmin$).
As we have mentioned,  for simplicity, we assume a simple box distribution for the energy resolution.
The results are very similar when using a more realistic distribution, such as Gaussian. 

For our figures we chose $m_\chi=$ 10 MeV, in  which case
$E'$ can go between 0.67~eV and 35.6~eV (when $\vmin=$ 800 km/s). For our plots we choose two representative values, $E'=5~{\rm eV}$ (considering that detecting at least one electron requires 2.9~eV in Ge and 3.6~eV in Si~\cite{Essig:2015cda}), and  $E'= 15$ eV.

We consider two DM-mediator form factors, $F_{\rm DM}=1$ for a heave mediator and $F_{\rm DM}= (\alpha m_e/q)^2$ for a light mediator, which generally appear in a  variety of models such as the scenarios of vector-portal DM with a dark photon mediator or magnetic-dipole-moment interactions.

For the dielectric function, we employ the output data from \textsf{DarkELF}~\cite{Knapen:2021bwg} and interpolate it into a continuous function. 
The dielectric function data are calculated using \textsf{GPAW}~\cite{Mortensen:2005,Enkovaara:2010}
that relies on first principles
time-dependent DFT. \textsf{DarkELF} allows the calculation to be carried out with or without considering screening  and local field effects (LFE) (i.e. including or excluding information on the non-diagonal components of the dielectric matrix).
In our Figs.~\ref{fig:dmflux} and \ref{fig:dmfluxscaled}   we show results  the following cases: 1) with screening and with LFE (red lines); 2) without screening effect (namely $|\epsilon|^2=1$) but with LFE (green lines); 3) without screening effect or LFE (blue lines). We also compare results from these three cases to results of \textsf{QEDark}~\cite{Essig:2015cda} (dark grey lines), as obtained in our previous paper~\cite{Chen:2021qao}.

It is worth noting that the maximum value  of $q$ for the \textsf{DarkELF} data is $22.5\,{\rm keV}$ while it is approximately $67\,{\rm keV}$ for \textsf{QEDark}. To compare the models for the same parameters range, we use $q_{\rm cut}=22.5\,{\rm keV}$ for all our  computations.
For $E'=5\,{\rm eV}$, we have $q_{\rm min}(E=4.5\,{\rm eV})=1.7\,{\rm keV}$, $q_{\rm max}(E=4.5 \,{\rm eV})=51.6 \,{\rm keV}$, and maximum $\tilde{q}(E=5.5\,{\rm eV})=10.5 \,{\rm keV}$. 
For $E'=15$~eV, $q_{\rm min}(E=13.5\,{\rm eV})=5.7\,{\rm keV}$, $q_{\rm max}(E=13.5\,{\rm eV})=47.7\,{\rm keV}$, and maximum $\tilde{q}(E=16.5\,{\rm eV})=18.2 \,{\rm keV}$. We can see that for the two representative energy values, we get the whole left branch and part of the right branch with this choice of momentum cut. The whole integral is then numerically evaluated over the range discussed above using \textsf{Mathematica}. 

\section{Results}

Figs.~\ref{fig:dmflux} and~\ref{fig:dmfluxscaled} show the response functions $\ud\mathcal{R}/\ud E'$ as a function of $\vmin$ in Ge, calculated with three different dielectric function computation methods using \textsf{DarkELF}, and additionally with \textsf{QEDark}. We find that the results for Si are very similar. They are shown in Figs.~\ref{fig:Sidmflux}  and~\ref{fig:Sidmfluxscaled}.

The plots are made with 100 $\vmin$ values, chosen with equal spacing over the domain $(0, 800)$ km/s. These figures show the response functions for detected energy $E'=5$ eV (left panels) and $E'=15$ eV (right panels), and two different DM form factors $F_{\rm DM}=1$ (upper panels) and $F_{\rm DM}= (\alpha m_e 1/q)^2$ (lower panels).

In Figs.~\ref{fig:dmflux} and~\ref{fig:Sidmflux}, the response functions are plotted in their original natural units. 
In Figs.~\ref{fig:dmfluxscaled} and~\ref{fig:Sidmfluxscaled} the response functions are instead scaled so that the maximum of each curve is close to 1 (which is the way the  \textsf{QEDark} results were shown in Ref.~\cite{Chen:2021qao}). Recall that the response function acts as a window function through which measured rates in direct detection experiments can provide information about the DM velocity (or speed) distribution through the function $\tilde{\eta}(\vmin)$. By scaling the response functions we can better appreciate the range of $\vmin$ selected by each of them. 

In Fig.~\ref{fig:dmflux} and~\ref{fig:Sidmflux} we see that the \textsf{QEDark} results (dark grey lines) in general are similar to the results using the unscreened without LFE \textsf{GPAW} calculation (blue lines), the unscreened with LFE \textsf{GPAW} calculation (green lines) is intermediate between the previous two and the result using the  screened with LFE \textsf{GPAW} calculation (red lines).
This is roughly in agreement with the results shown in Fig.~5 of Ref.~\cite{Knapen:2021run} for upper limits on $\sigma_{\rm ref}$ as function of $m_\chi$ for the four different models.
There too the \textsf{QEDark} results are similar with the unscreened without LFE limits and they the least restrictive, the unscreened with LFE limits are intermediate and the most restrictive limits are those of the  screened with LFE \textsf{GPAW} calculation.

We can also see that the screening effect effectively reduce the amplitude of the response function, and this effect is much more pronounced for $F_{\rm DM}\sim 1/q^2$, especially in the low $E'$ regime. 

The weight assigned by the response functions calculated in different cases as window function to different values of $\vmin$ are in general very similar, as we can see in Figs.~\ref{fig:dmfluxscaled} and~\ref{fig:Sidmfluxscaled}. However,  the shape of the window function can change considerably in the regime where the screening effect is much more pronounced, as shown in the lower left panel of Fig.~\ref{fig:dmfluxscaled} and~\ref{fig:Sidmfluxscaled}.
 
\section{Conclusions}

Light sub-GeV DM constitutes a promising target for exploration in direct DM detection experiments studying DM-electron scattering. While conventional direct DM detection analysis depends on assumptions about poorly known local halo DM distribution, the halo-independent analysis allows to infer local DM halo properties from direct detection data and consistently analyze distinct experimental targets. In crystal target materials, which allow achieving lower experimental thresholds, collective in-medium effects could modify signatures of DM interactions. Here we formulate for the first time the methodology for halo-independent direct DM detection analysis for DM-electron scattering including in-medium effects. We show for a germanium target, and similarly for a silicon target, that in-medium effects could significantly impact the interpretation of direct DM detection data. Thus, 
such effects must be included in future halo-independent analyses for proper inference of local DM halo properties from DM detection 
data.

\section*{Acknowledgments}

\vspace{-1em}
We thank Tongyan Lin for comments and Tien-Tien Yu for discussions and clarifications regarding the \textsf{QEdark} package.
The work of G.B.G. and M.C. was supported in part by the U.S. Department of Energy (DOE) Grant No. DE-SC0009937. V.T. was also supported by the World Premier International Research Center Initiative (WPI), MEXT, Japan.

\bibliography{biblio.bib}
\end{document}